# Magnetic glassy state at low spin state of Co$^{3+}$ in EuBaCo$_2$O$_{5+\delta}$ ($\delta$ = 0.47) cobaltite

Archana Kumari[1*], C. Dhanasekhar[1,2], Praveen Chaddah[3†], D Chandrasekhar Kakarla[4,5], H. D. Yang[4,5], Z. H. Yang[4], B. H. Chen[6], Y. C. Chung[6] and A. K. Das[1]

[1]Department of Physics, Indian Institute of Technology, Kharagpur 721302, West Bengal, India

[2]Department of Physics, Indian Institute of Technology Bombay, Powai, Mumbai 400076, India

[3]UGC-DAE Consortium for Scientific Research, Khandwa Road, Indore 452001, Madhya Pradesh, India

[4] Department of Physics, National Sun Yat-Sen University, Kaohsiung 804, Taiwan

[5]Center of Crystal Research, National Sun Yat-Sen University, Kaohsiung 804, Taiwan

[6]National Synchrotron Radiation Research Center, Hsinchu, 30076, Taiwan

[*]tiwariarchana9@phy.iitkgp.ernet.in

[†] Since retired; chaddah.praveen@gmail.com

## ABSTRACT

The magnetic glassy state is a fascinating phenomenon, which results from the kinetic arrest of the first order magnetic phase transition. Interesting properties, such as metastable magnetization and nonequilibrium magnetic phases, are naturally developed in the magnetic glassy state. Here, we report magnetic glass property in the low spin state of Co$^{3+}$ in EuBaCo$_2$O$_{5+\delta}$ ($\delta$ = 0.47) cobaltite at low temperature ($T < 60$ K). The measurements of magnetization under the cooling and heating in unequal fields, magnetization relaxation and thermal cycling of magnetization show the kinetic arrest of low magnetization state below 60 K. The kinetically arrested low temperature magnetic phase is further supported through the study of isothermal magnetic entropy, which shows the significant entropy change. The present results will open a new window to search the microscopic relation between the spin state transitions and the kinetic arrest induced magnetic glassy phenomena in complex materials.

## 1. Introduction

The first-order magnetic phase transition (FOMPT) in magnetic materials shows several fascinating phenomena, such as giant magnetocaloric effect, colossal (or giant) magnetoresistance, magnetic shape memory effect, etc [1–5]. In recent years, the coexistence of magnetic phases (or multiple magnetic phases) is observed in some magnetic materials below the FOMPT. The phase coexistence arises from the kinetic arrest of the supercooled, metastable, and nonequilibrium magnetic phase in the background of a stable equilibrium magnetic phase. Under certain conditions, the nonequilibrium supercooled phase is transformed into the stable equilibrium phase and the nonequilibrium phase is named as magnetic glass (MG) [5–12]. The origin of MG phenomenon is different from the conventional spin glass (SG) and the reentrant spin glass (RSG) and is observed in a variety of magnetic systems [13,14]. In the majority of MG materials, it is observed that the FOMPT is accompanied with thermomagnetic irreversibility (TMI) and this TMI can be understood from the appearance of phase coexistence and metastability [4–13].

Further, the kinetically arrested FOMPT is coupled with a structural change in most of the MG materials. For examples, the high temperature ferromagnetic (FM) to low temperature antiferromagnetic (AFM) transition in chemically doped-$CeFe_2$ alloys [6,15,16] involves cubic to rhombohedral structural changes, whereas the high temperature AFM to low temperature FM transition in $Gd_5Ge_4$ [17] is associated with the orthorhombic structural transition from the $Sm_5Ge_4$-type to orthorhombic $Gd_5Si_4$-type. On the other hand, the formation of MG is also reported in the layered $YBaCo_2O_{5+\delta}$ ($\delta = 0.50$) cobaltite with 5% Ca substitution at Y and Ba sites [10]. In complement to the Ca substitution studies, we have shown that the electron doped $YBaCo_2O_{5.5-\delta}$ ($\delta = 0.13$) cobaltite also showed the presence of MG state at low temperature ($T$) [18]. However, magneto structural coupling hardly plays a role in both the cobaltites and the MG phenomenon is explained considering the local phase separation of FM and AFM clusters.

Our recent studies [19] on layered $EuBaCo_2O_{5+\delta}$ ($\delta = 0.47$) (EBCO) cobaltite showed metal to insulator transition at 355 K ($T_{MIT}$), paramagnet (PM) to FM transition at 267 K ($T_C$), FM to AFM1 transition at 240 K ($T_{N1}$) and AFM1–AFM2 transition at 204 K ($T_{N2}$). The FM to AFM transition showed the significant first order thermomagnetic hysteresis effect and further, we showed that the AFM phases were shifted to low temperature under the external magnetic field. More importantly, the octahedral $Co^{3+}$ showed the high spin (HS) to intermediate spin (IS) state transition at $T_{MIT}$ and this IS state gradually converted to the low spin (LS) state below 155 K.

In this article, we report the kinetic arrest of magnetization at low temperature in the presence of the magnetic field (H) and a glasslike dynamical response in $EuBaCo_2O_{5+\delta}$ ($\delta = 0.47$) (EBCO) cobaltite. The non-equilibrium magnetic glassy state in EBCO also contributes a significant entropy change at low temperature. The observed MG phenomena in this cobaltite differs from the other magnetic glassy materials because the various high temperature magnetic phases (FM, AFM1 and AFM2) in EBCO are strongly coupled with the $Co^{3+}$ spin state transitions and the kinetic arrest is observed at the low temperature, where LS (S=0) state of $Co^{3+}$ presents. Finally, we show that the kinetic arrest of FOMPT in EBCO is not associated with a structural transition, and the other various possibilities that can elucidate the origin of kinetic arrest and MG phenomena are discussed.

## 2. Experiment details

The polycrystalline EuBaCo$_2$O$_{5+\delta}$ ($\delta = 0.47$) cobaltite was prepared by conventional solid-state reaction method and the details can be found in Ref. 19. The temperature-dependent structural analyses were performed using the synchrotron X-ray powder diffraction (SXRPD) at TPS-09A beam line at the National Synchrotron Radiation Research Centre, Hsinchu, Taiwan. Finely grounded powder sample was loaded in 0.3 mm glass capillary. The diffraction patterns were measured in the range of $2\theta = 1.8–100°$ with a step size of $\Delta(2\theta) = 0.00375°$ with the incident wave length of 0.61992 Å. The magnetization versus temperature, i.e. M ($T$) measurements were carried out using Physical Property Measurement System (PPMS), Cryogenic Limited, UK. AC magnetization, time dependence of magnetization i.e. M (t) and temperature cycling magnetization measurements were performed using Magnetic Property Measurement System (MPMS SQUID VSM), Quantum Design, USA.

## 3. Results

### 3.1. Ac magnetization studies

To confirm the various dc M transitions and to find out whether presence of any conventional spin glass signature in EBCO, a frequency dependent ac M measurements under an ac field of 1 Oe were performed and the obtained real part of $\chi'_{ac}$ ($T$) is shown in figure 1. The various dc M transitions [19] also appear in $\chi'(T)$ and those transitions are frequency independent in the measured frequency range. The $\chi'_{ac}$ ($T$) curves show a drop below 60 K, which matches with the zero field cooled (ZFC) M behavior (see figure 2(b)). Further, to ruled out the spin or cluster glass like behavior below 60 K, we performed a memory effect in $\chi'_{ac}$ ($T$) and the result is shown in the inset of figure 1. In these measurements, the sample was cooled to the base temperature of 5 K while administering halt at 25 K for $t_w$ = 3 hours. After reaching 5 K, the sample was warmed to 380 K and $\chi'_{ac}$ ($T$) was recorded during the heating cycle and the result is shown in the inset of figure 1. The data taken after the halt is referred as $\chi'_{ac}$ $halt$ and referred as $\chi'_{ac}$ $ref$ without the halt time. The difference between the $\chi'_{ac}$ $halt - \chi'_{ac}$ $ref$ ($\approx \Delta\chi'_{ac}$) is also shown in the inset of figure 1. In most of the spin glass or cluster glass the $\Delta\chi'_{ac}$ showed dip kind of signatures at the halt temperatures [20] and in our present case the $\Delta\chi'_{ac}$ have not shown any such signatures at 25 K. The frequency independent $\chi'_{ac}$ ($T$) peaks in combined with $\Delta\chi'_{ac}$ result ruled out the SG and cluster glass (CG) like features below 60 K. The observed ac magnetization results are in line with the single crystal studies of GdBaCo$_2$O$_{5+x}$, where it is showed that the conventional spin-glass behavior is not possible for any x and suggests that all magnetic irreversibilities are interrelated to the spin rearrangement within a spin-ordered state [21].

### 3.2. Kinetic arrest and metastability

The detailed ac M studies show the absence of conventional glass like feature in the EBCO and the detailed dc M studies show the signatures of phases coexistence [19]. To identify the kinetic arrest-induced phase coexistence, Banerjee et al. proposed a special magnetic measurement protocol, i.e., cooling and heating in unequal fields ("CHUF") [22,23]. Following this protocol, the EBCO sample was cooled from the 380 to 10 K under various cooling magnetic fields (H$_{cool}$), i.e. 0, 1, 2, 3, 5 and 7 T respectively and at 10 K the H$_{cool}$ is changed to measuring magnetic field (H$_{measure}$) of 2 T and M was reordered on warming cycle (see figure 2(a)). For $T > T_{N1}$ the M behavior of EBCO under the CHUF protocol remains the same irrespective of the H$_{cool}$. For $T < T_{N1}$, the M curves show strong dispersion and more interesting behavior is observed below 60 K. For $H_{cool} = 0$ T, the M of the sample shows an initial increase from 5-60 K and becomes flat. Whereas in case of cooling under 1 T and measuring at 2 T the M increases up to 35 K and becomes flat or plateau is seen. While for $H_{cool} = H_{measure}$ the M follows the same path without showing any initial rise. For $H_{cool} > H_{measure}$ the initial increase in M is absent and shows slightly decrease in M with increasing the $H_{cool}$. This result clearly suggests that the low temperature M is metastable and is kinetically arrested while cooling. Further, the kinetically arrested M state is de-arrested and reaches a stable magnetization above 60 K. We note that the de-arrest takes place at low $T$ with increasing the cooling H. This is further supported by the ZFC measured at different low fields, where the initial increase in M is shifted to low $T$ with increasing the H as shown in the figure 2(b). The H-T phase diagram along with kinetic arrest band obtained from the ZFC magnetization is shown in the figure 2(d).

The M relaxation measurements at 20 K are measured under 2 T for the ZFC and the field cooled (FC) cases on EBCO provide further evidence for the metastability of the lower M state. In ZFC measurement, the EBCO sample was cooled from 380 K to 20 K in zero field; then the sample was held at $T = 20$ K for a time $t_w = 300$ s; after then H was applied and M was recorded as a function of time. In FC measurement the sample was cooled from 380 K to 20 K in presence of an applied magnetic field (H); then the sample was held at $T = 20$ K for a time $t_w = 300$ s; after which the H was removed and the M was recorded as a function of time. The ZFC magnetization of EBCO is increased from 0.102 to 0.110 $\mu_B$/f.u. and the corresponding change in M is ~ 7.8 % (figure 2(c)). On the other side, the FC magnetization is decreased from 0.176 to 0.172 $\mu_B$/f.u. and the corresponding change in M is ~ 2.2 % (figure 2(c)). At 20 K the initial FC magnetization is almost double of the ZFC magnetization, which indicates that the FC magnetization of EBCO is close to the equilibrium state, whereas the ZFC magnetization has very low value and is metastable and relaxes fast.

To understand the metastable nature of the low temperature ZFC magnetization of EBCO, we have further measured the effect of $T$ cycling on M applying H of 3 T and the obtained results along with the field cooled cooling (FCC) and field cooled warming (FCW) magnetization curves are shown in figure 3. For better understanding, the reference ZFC magnetization of the EBCO obtained under H of 3 T is also shown in the inset of Fig. 3. In ZFC temperature cycling, initially the sample was cooled from 380 K to 10 K and at 10 K, the magnetic field of 3 T was applied and M was recorded while increasing as well as decreasing the temperature in step of 10 K from 10 K to 120 K. In these measurements, the M value increases continuously from that of the previous one and above 60 K the ZFC magnetization matches with the FCW magnetization. This observation is very similar to that of the reported one for Gd$_5$Ge$_4$ [24]. This clearly suggests that, at low temperature, even at a fixed H, the thermal cycling is able to convert some of the supercooled metastable low M phase to equilibrium higher M phase. This de-arrest of the metastable state is occurred over a range of temperature which is identified as a 'kinetic arrest band'. Similar result is obtained under the ZFC $T$ cycling at H = 5 T, which implies that the observed metastable nature persists even at high H (see inset of figure 3). The temperature over which this de-arrest is occurred or the 'kinetic arrest band' is shifted to lower temperature at larger H.

### 3.3. Magnetic entropy studies

To further investigate the MG property and the nature of the various magnetic phases, the magnetocaloric (MC) measurement was performed on EBCO. Usually, most of the MC studies are devoted to understand the usefulness of the materials for refrigeration applications; however, this technique can also be used to probe the H induced magnetic phase transitions. The $T$ variation of isothermal magnetic entropy change ($\Delta S_M$) has been calculated from the isothermal M curves using Maxwell's equation

$$|\Delta S_M| = \mu_o \int_0^{H_{max}} \left(\frac{\partial M}{\partial T}\right)_H dH$$

The above relation states that the change in entropy of any material is directly associated with the first derivative of M with respect to $T$, which makes it more sensitive to probe the small change in M. In most of the studies, the $T$ variation of entropy change is represented with -$\Delta S_M$ vs. $T$ curve, which shows a positive peak in the vicinity of the FM transition and a negative peak at the onset of the AFM ordering.

The obtained -$\Delta S_M$ as a function of $T$ for different $\Delta H$ for the EBCO is shown in figure 4. We have noticed a positive symmetric peak in -$\Delta S_M$ curves at $T_C$ and a crossover from positive to negative in -$\Delta S_M$ curves at $T_{N1}$, which is highlighted in the inset of figure 4. While at $T_{N2}$ the -$\Delta S_M$ curves show the broad asymmetric negative peak. More importantly, the positive to negative crossover and the position of the maximum in -$\Delta S_M$ curves at $T_{N1}$ and $T_{N2}$ are found to be shifted towards the low $T$ with increasing H, which is correlated with the M ($T$) behaviour [19]. On the other side the peak position of -$\Delta S_M$ curve at $T_C$ is insensitive to H. The -$\Delta S_M$ curves show the maximum value of 0.67 J/kg K (-0.35 J/kg K) at $T_C$ ($T_{N2}$) for the maximum H of 7 T. Interestingly, -$\Delta S_M$ increases linearly with H when T decreases for $T < 60$ K and at 10 K it is found to be -0.61 J/kg K for H = 7 T. The total magnetic entropy change due to the spin only is related to $\Delta S_M$ $\alpha$ $\ln(2S + 1)$, where S is the total spin. According to this relation one would expect $\Delta S_M$ ((IS); S =1) > $\Delta S_M$ ((LS); S =0). The obtained $\Delta S_M$ values at 7 T are far below than the theoretical spin only values. However, the observed magnitude of $\Delta S_M$ gives the additional information about the Co$^{3+}$ spin state transitions. In the vicinity of AFM2, the observed $|\Delta S_M|$ at 7 T is almost half of the FM phase. As discussed in ref. 19 that in AFM2 phase, the IS states of Co$^{3+}$ is gradually converted to LS state and the complete LS state of Co$^{3+}$ would be expected at low $T$. At low $T$ one would expect that $|\Delta S_M|$ will tend to be constant (if one considers the contribution of pyramidal Co$^{3+}$ IS state). On contrary $|\Delta S_M|$ is increased with H below 60 K, where the magnetic glassy phase is observed. The magnetic entropy studies clearly show that the arrested low $T$ magnetic phase contributes significant entropy change.

## 4. Discussion and concluding remarks

The comprehensive measurements of M on EBCO provide a macroscopic evidence of the kinetic arrest and the formation of magnetic glassy state below 60 K. As discussed in the introduction, the magneto-structural transitions are the major ingredients that drive the FOMPT and cause the kinetic arrest in most of the magnetic glassy materials. To understand the effect of structural changes in various magnetic transitions, we have reordered the $T$ variation of synchrotron X-ray diffraction (SXRPD) on EBCO sample in the temperature range of 99 - 400 K. The SXRPD patterns match with the other members of RBaCo$_2$O$_{5+x}$ (x ≈ 0.5) cobaltites [25–29] in which the crystal symmetry was found to be invariant, i.e. orthorhombic (*Pmmm*) in the whole investigated $T$ range. figures 5(a & c) show the SXRPD pattern of selected 2θ range for the 400 K and 99 K respectively. figure 5(b) shows the contour map of intensity variation for the selected 2θ range with respect to $T$. The main three intensity reflections, (120), (022) and (102), corresponding to the *Pmmm* space group have been shifted to the higher angle with decrease of $T$. However, this trend was found to be dramatic in the vicinity of $T_{MIT}$ (350 K-310 K), where reflections (120) and (102) are shifted towards the lower angle and (022) is shifted to the higher angle. Figure 5(d) shows the $T$ dependence of the lattice constants of EBCO, where the lattice parameters show unusual variation in the vicinity of $T_{MIT}$ and HS to IS transition of $Co^{3+}$ is observed. With decreasing of $T$ in the range of 360-300 K, both a & c parameters show negative drops and at the same $T$, b parameter shows positive jump. Further decreasing of $T$, such anomalous change is not observed but gradual decrease is found in the lattice parameters. The anomalous changes in lattice constants close to $T_{MIT}$ matches with the results of the latent heat measured by differential scanning calorimetry (DSC), which indicates that the transition is of first-order type [19]. This overall behavior of the present EBCO matches with the other members of this family where the first order structural changes are coupled with the metal to insulator transition ($T_{MIT}$) and also with the spin state transitions of $Co^{3+}$.

Further, the thermal studies on different cobaltites of this family show that the PM–FM transition is second order in nature and FM–AFM1 is the first order magnetic transition [30]. Although the zero field SXRPD studies do not show any first order signatures at FM–AFM1 transition ($T_{N1}$=240 K) in EBCO but the M ($T$) and M (H) studies show that there is significant thermal hysteresis, which confirms that the $T_{N1}$ is the first order in nature. For instance, metamagnetic transitions are also observed in the EBCO, and more importantly, the metamagnetic transitions are shifted to the high field with decreasing $T$ and are found to be incomplete at low $T$ up to 7 T [19,31]. Such behavior has been observed in a variety of magnetic materials, and in these materials, the kinetic arrest arises due to the incomplete first-order phase transition [8,23,32].

Compared to the existing magnetic glassy materials, the MG state in EBCO seems unusual. As discussed above that the first order structural changes seem almost decoupled with the magnetic transitions and more importantly the magnetic active ion i.e. $Co^{3+}$ shows $T$ variation of the spin state transitions. In EBCO, at $T_{MIT}$ the octahedral $Co^{3+}$ shows HS to IS state transition and with decreasing $T$ the IS of $Co^{3+}$ is gradually transformed into LS state [19]. The muon-spin relaxation measurements on the $RBaCo_2O_{5.5}$ cobaltites showed that the AFM1 and AFM2 phases have different types of spin state order (SSO) [33]. The AFM1 state is having two types of SSO, while AFM2 phase is having up to four different types of SSO structures. In both AFM1 and AFM2 phases, the IS $Co^{3+}$ ions on pyramidal sites show AFM coupling, while the neighboring octahedral $Co^{3+}$ holds either IS or LS state. Although the AFM1 and AFM2 phases having the different types of the SSO structures, but the self-energies of these SSO structures are almost same and causes the microscopic phase separation at FM-AFM1 transitions [33]. Further, it was suggested that these two AFM phases were independent of each other and developed as a well separated phases with decreasing $T$. This has been further confirmed experimentally in our previous studies of EBCO cobaltite through the M ($T$) and M (H) studies [19].

Here, we further argue that the identification of non-equilibrium magnetic phase at low $T$ in the EBCO is very difficult; because the octahedral HS $Co^{3+}$ (S=2) sites are transformed into the diamagnetic LS $Co^{3+}$ (S = 0) with the decreasing $T$, and the pyramidal IS $Co^{3+}$ sites have an AFM coupling. One would expect a very low magnetic moment at low $T$. As shown in figure 2 (a), the change of the magnetic moment between 60 -10 K in the arrested $T$ region is very small, i.e., 0.2 μ$_B$/f.u. In most of the reported magnetic glassy materials, the higher $T$ magnetic phase is generally arrested and shows a non-equilibrium glassy signature at low $T$. Ideally, in EBCO, the high $T$ magnetic phase, i.e. ferri or ferromagnetic phase has to be kinetically arrested at low $T$ in the background of the AFM phase and would expect to show a non-equilibrium glassy signature. In contrast, the obtained kinetic arrest band (figure 2(d)) at different magnetic fields matches with the ideal FM ground states (the high $T$ magnetic phase is AFM) [34]. Further, the small kinetically arrested magnetic moment at low $T$ may be due to several factors, such as partially arrested high $T$ ferro or ferrimagnetic phase, or incomplete first order metamagnetic phases or from one of the arrested SSO ordered AFM phase. These signatures clearly indicate the complex behavior of the arrested phase.

Finally, it is worth to compare the MG state of the present EBCO with the other available reports of this cobaltite family. The MG state is reported in the $YBaCo_2O_{5+\delta}$ ($\delta = 0.50$) cobaltite with 5% Ca substitution at Y and Ba sites [10] and also in the electron doped $YBaCo_2O_{5.5-\delta}$ ($\delta = 0.13$) [18] cobaltite. Although the MG state is observed for all samples at low $T$, but the high $T$ magnetic state is very different in the case of EBCO as compared to the both Y cobaltites. In Y cobaltites, the AFM2 phase, field induced multiple metamagnetic transitions and the signatures of octahedral $Co^{3+}$ IS to LS state transition are not observed. The evaluation of phase separation between the FM and AFM clusters in the doped Y cobaltites causes the MG state at low $T$. It is noted that the phase separation in the case of undoped $YBaCo_2O_{5+\delta}$ ($\delta = 0.50$) is minimal and sufficiently significant in case of $EBaCo_2O_{5+\delta}$ ($\delta = 0.47$) [19]. This difference clearly indicates that the microscopic origin of MG state at low $T$ in the present EBCO is different from the doped Y cobaltites, where the intrinsic magnetic phase separation in EBCO is coupled with the octahedral $Co^{3+}$ spin state transitions.

In conclusions, we have firmly established the formation of the magnetic glassy state in $EuBaCo_2O_{5+\delta}$ ($\delta=0.47$) cobaltite through various magnetization measurements. The nonequilibrium magnetic glassy state is observed at the low spin state of $Co^{3+}$, where the arrested phase has very low magnetization. The isothermal magnetic entropy studies also show the large entropy change at the kinetically arrested phase. Though this study demonstrates a possible mutual coupling among the magnetic phase separation, spin state transition and the magnetic glassy state, but the microscopic origin of the kinetic arrest that has led to the glassy dynamics in EBCO is still lacking and challenging. Further, the observed MG state in this cobaltite family is generic, not limited to the above studied materials but also can be extended to various chemical substitutions as well as for the nanostructured thin films. The structural and magnetic imaging techniques in the presence of magnetic field may play a key role in understanding the microscopic relationship between the kinetic arrest and the supercooling phase [35–38]. Such studies should be initiated to a large extent to investigate the microscopic relationship between kinetic arrest and the spin state transition driven magnetic phase separation in these cobaltites.

### Acknowledgements

The authors of IIT Kharagpur acknowledge IIT Kharagpur funded VSM SQUID magnetometer at central research facility and cryogenic physical property measurement system at department of physics. Archana Kumari acknowledges the MHRD, India for providing the senior research fellowship.

**Figure captions:**

**Figure 1.** (Color online) In-phase ac-magnetization measured at different frequencies as a function of temperature for EBCO sample. Insets show the different magnetic transitions and memory effect.

**Figure 2.** (Color online) Temperature dependent magnetization obtained for EBCO sample (a) during warming after being cooled in different magnetic fields and (b) in ZFC mode under various applied fields. (c) Magnetization versus time plot for EBCO at 20 K along ZFC and FC path in the presence of H = 2 T. (d) Schematics of kinetic arrest band diagram as the corresponding ZFC magnetization curves for the case with high magnetization ground state.

**Figure 3.** (Color online) Magnetization versus temperature plots obtained in ZFC, FCC, and FCW modes in an applied field of 30 kOe for EBCO sample. In the ZFC mode the sample is subjected to thermal cycling at various temperatures and the detailed measured protocol is described in the corresponding text. The top inset shows the ZFC mode thermal cycling of magnetization in an applied field of 50 kOe. The lower inset shows the ZFC magnetization without thermal cycling, i.e. normal ZFC magnetization.

**Figure 4.** (Color online) Plot of the change in magnetic entropy ($-\Delta S_M$) for 1 T field interval as a function of temperature for EBCO sample calculated using the thermodynamic Maxwell relation. Insets show the expanded view of $-\Delta S_M$ around the $T_{N1}$ (240 K).

**Figure 5.** (Color online) Selected portion of the SXRPD patterns ($2\theta = 12.87°$-$13.24°$) at (a) 400 K and (c) 99 K respectively. (b) Contour map of three main reflections for the selected $2\theta$ (= $12.87°$-$13.24°$) for the temperature range of 400 - 99 K; the intensity variation is denoted by the colour profile and the reflections are indexed on the basis of orthorhombic (*Pmmm*, $a_p$ x $2a_p$ x $2a_p$) lattice. (d) Temperature dependence of lattice constants of $EuBaCo_2O_{5.47}$.

**Figures:**

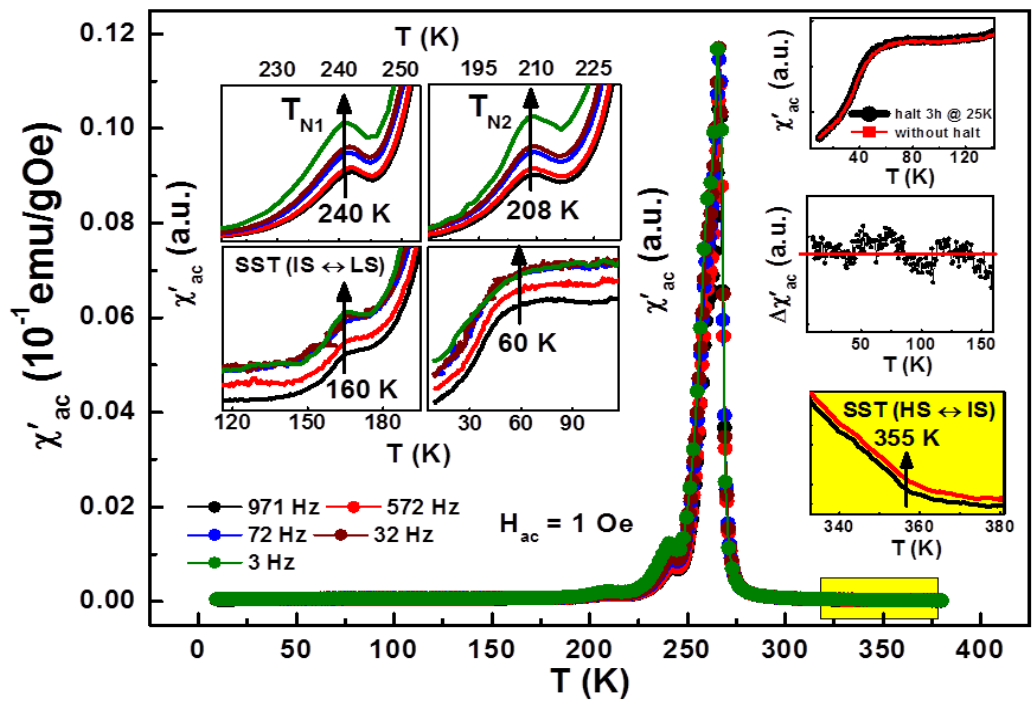

**Figure 1.**

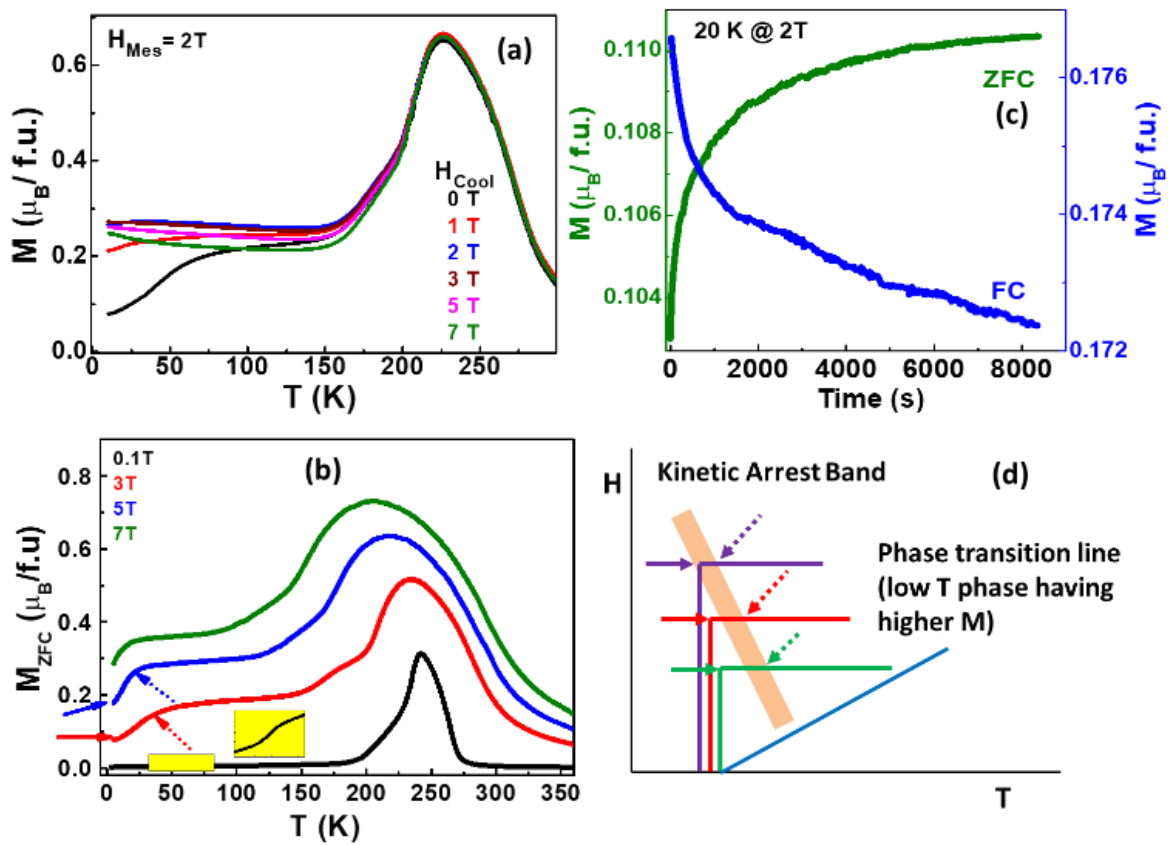

**Figure 2.**

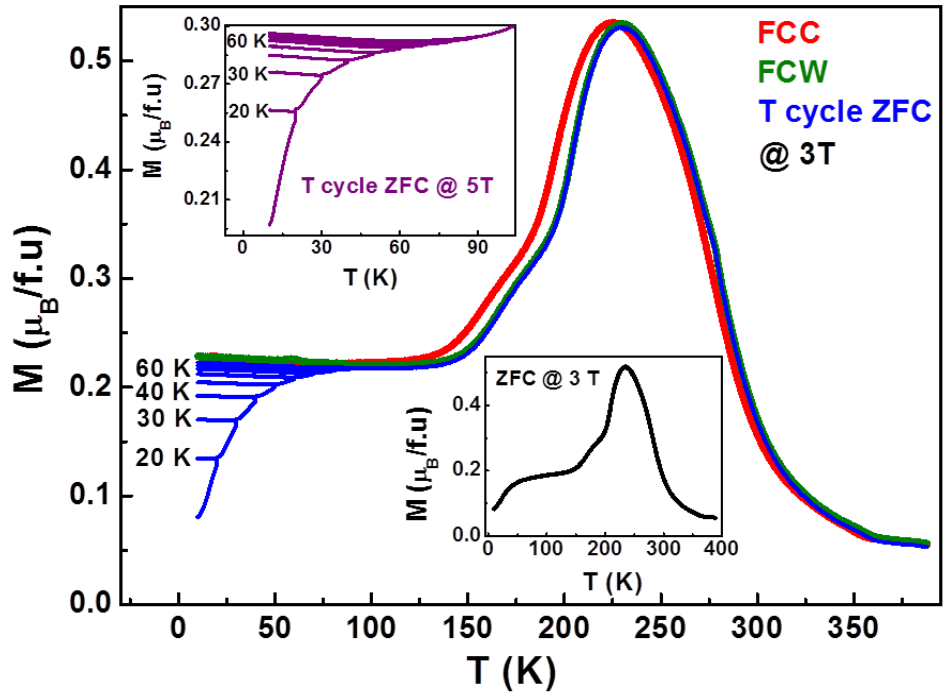

**Figure 3.**

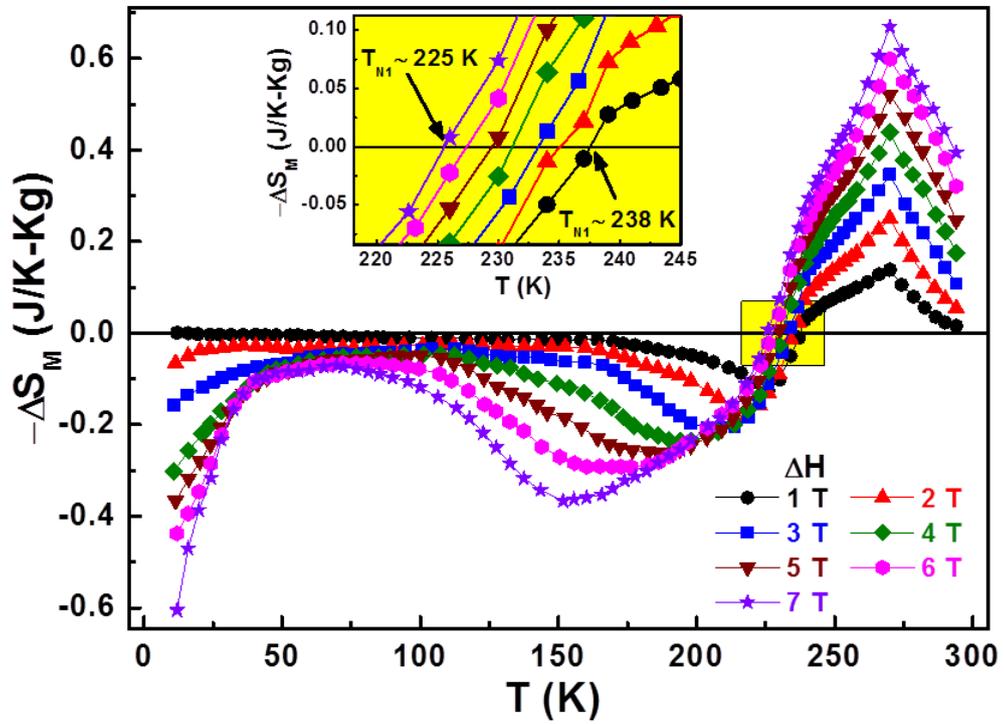

**Figure 4.**

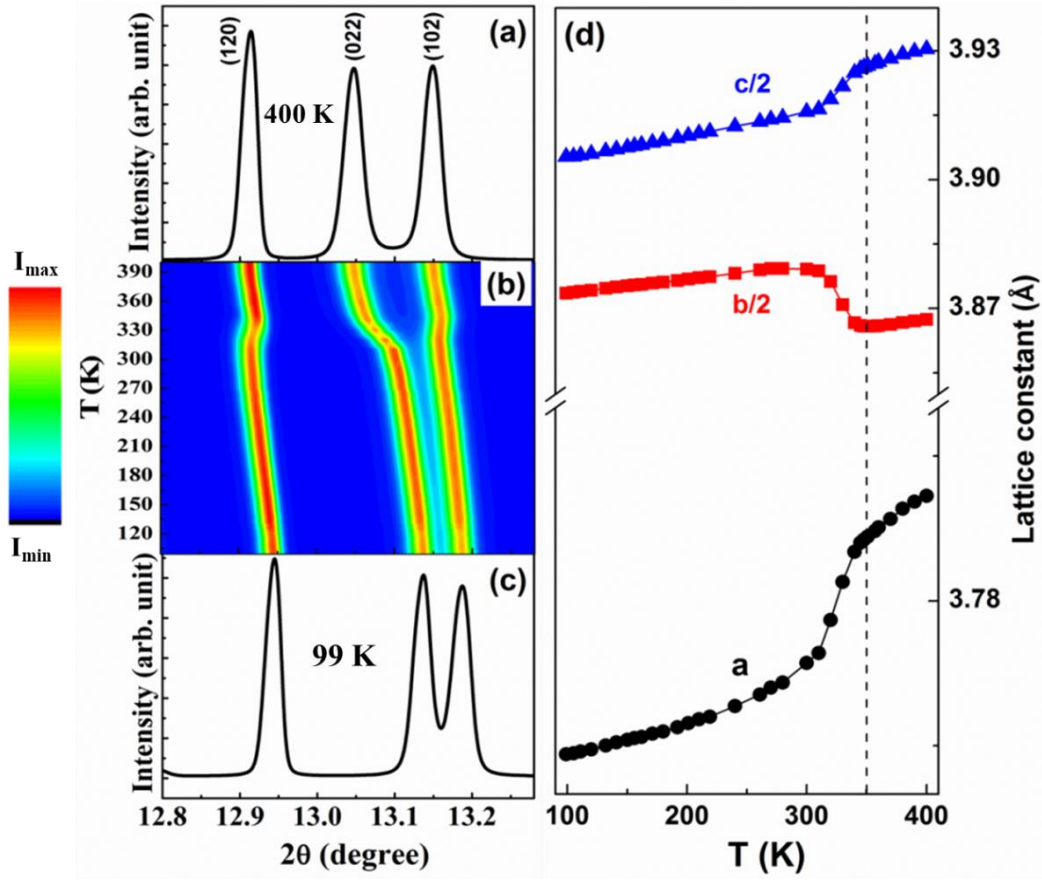

Figure 5.